\documentclass[cameraready]{Interspeech}

\title{Rethinking Speech Foundation Model Fine-tuning: \\
Better SFT or Better Match?}

\author[affiliation={1}]{Wangjin}{Zhou}
\author[affiliation={1}]{Yizhou}{Zhang}
\author[affiliation={1}]{Yichi}{Wang}
\author[affiliation={1}]{Tatsuya}{Kawahara}

\address{
    $^1$ Graduate School of Informatics, Kyoto University, Kyoto, Japan 
}

\email{\{zhou\}@sap.ist.i.kyoto-u.ac.jp}


\usepackage{comment}
\usepackage{amsmath}
\usepackage{amsfonts}
\usepackage{amssymb}
\usepackage{booktabs}
\usepackage{array}
\usepackage{diagbox}
\usepackage{makecell}
\usepackage{graphicx} 
\usepackage{multirow}
\usepackage{makecell}
\usepackage{tabularx}
\usepackage{array}
\usepackage{pifont}
\usepackage{xurl}
\usepackage{hyperref}
\usepackage{adjustbox}

\newcolumntype{L}{>{\raggedright\arraybackslash}X}
\newcolumntype{C}{>{\centering\arraybackslash}X}

\begin{document}

\maketitle

\begin{abstract}
Supervised fine-tuning (SFT) is widely used to adapt self-supervised speech representations to downstream classification tasks. Small gains observed under a single pretrained checkpoint are often interpreted as method-level improvements, i.e., a higher attainable performance ceiling. We show that such conclusions are not always reliable because SFT outcomes depend strongly on the specific pretrained instance. We conduct a systematic study on 3 SUPERB classification tasks, evaluating 8 SFT variants across 9 pretrained checkpoints from wav2vec~2.0, HuBERT, and WavLM, with multi-seed repetitions on representative base-scale models. We find that the identity of the statistically indistinguishable top-group SFT recipe is often checkpoint-dependent, with limited transferability across pretrained instances. These findings suggest that many reported downstream gains reflect instance and seed dependent \textit{elicitation match}, rather than universally improving the attainable performance ceiling.

\end{abstract}

\section{Introduction}

Supervised fine-tuning (SFT) is a dominant paradigm for adapting self-supervised pretrained speech models \cite{baevski2020wav2vec, hsu2021hubert, chen2022wavlm} to downstream classification tasks \cite{gao2023two, gao2024enhancing, 10446041}. In common practice, researchers compare SFT variants by downstream performance and often treat small gains under a single pretrained checkpoint as evidence that a recipe is ``better,'' i.e., achieving a higher attainable performance ceiling \cite{10848773, 10889634}.

This practice implicitly assumes that the relative effectiveness of SFT methods is stable across pretrained model instances. In other words, a recipe that is ``better'' under one pretrained checkpoint is expected to remain competitive under other checkpoints of comparable capability \cite{11353442, sang2024efficient}. However, this assumption overlooks an important interaction in the adaptation process. Downstream performance is jointly shaped by the backbone architecture, the pretraining data, and the SFT recipe \cite{Zhang2025SONARSC}. Yet existing evaluations often consider these factors in isolation. In particular, the interaction between a recipe and a specific pretrained instance is rarely examined \cite{merchant-etal-2020-happens, Zhang2025OnTI}. As a result, even when the architecture and the model size are held fixed, the apparent superiority of an SFT method may depend on which pretrained checkpoint it is applied to.

We propose to view SFT as a process of \emph{capacity elicitation}: fine-tuning attempts to activate and reorganize latent knowledge already present in pretrained representations. Under this view, apparent SFT superiority can reflect \emph{elicitation match}, meaning that a recipe more reliably activates the checkpoint's usable capacity rather than increasing the capacity itself. Different SFT configurations impose different elicitation biases, which may match some pretrained instances better than others, leading to differences in activation reliability. Consequently, two SFT methods can exhibit ranking discrepancies across checkpoints: Method~A may outperform Method~B on one pretrained instance, while Method~B surpasses Method~A on another. Such discrepancies challenge the external validity of conclusions drawn from single-checkpoint comparisons.

\begin{table*}[t]
\centering
\fontsize{5.6}{6.0}\selectfont
\setlength{\tabcolsep}{1.9pt}
\renewcommand{\arraystretch}{1.2}
\caption{Pretrained SSL checkpoints used in our study, with model size and pretraining data.}
\label{tab:ssl_models}
\scriptsize
\resizebox{\textwidth}{!}{
\begin{tabular}{l l c l}
\hline
\textbf{Checkpoint} & \textbf{Family} & \textbf{\#Params (M)} & \textbf{Pretraining data} \\
\hline
\texttt{facebook/wav2vec2-base-960h}            & wav2vec 2.0 & 95.00  & LibriSpeech 960h \\
\texttt{facebook/wav2vec2-base-100k-voxpopuli}  & wav2vec 2.0 & 95.00  & VoxPopuli (100k unlabeled subset) \\
\texttt{facebook/wav2vec2-large-960h}           & wav2vec 2.0 & 317.00 & LibriSpeech 960h \\
\texttt{facebook/wav2vec2-large-100k-voxpopuli} & wav2vec 2.0 & 317.00 & VoxPopuli (100k unlabeled subset) \\
\hline
\texttt{facebook/hubert-base-ls960}             & HuBERT      & 95.00  & LibriSpeech 960h \\
\texttt{facebook/hubert-large-ll60k}            & HuBERT      & 317.00 & Libri-Light 60k \\
\hline
\texttt{microsoft/wavlm-base}                   & WavLM       & 94.70  & LibriSpeech 960h \\
\texttt{microsoft/wavlm-base-plus}              & WavLM       & 94.70  & Libri-Light 60k + GigaSpeech 10k + VoxPopuli 24k \\
\texttt{microsoft/wavlm-large}                  & WavLM       & 316.62 & Libri-Light 60k + GigaSpeech 10k + VoxPopuli 24k \\
\hline
\end{tabular}
}
\end{table*}

To study this effect, we conduct a large-scale empirical investigation of SFT behavior in downstream speech classification. We evaluate eight SFT configurations along two axes: which intermediate representations are used for supervision, and how much of the pretrained network is frozen during adaptation. These configurations are applied to nine widely used self-supervised checkpoints from wav2vec~2.0, HuBERT, and WavLM. Experiments are conducted on three SUPERB classification tasks \cite{yang2021superb}: intent classification (IC) \cite{Tian2020ImprovingES, 9413388}, emotion recognition (ER) \cite{11270007, 9747870}, and speaker identification (SID) \cite{10094954, 11075516}. To characterize run-to-run variability, we further repeat experiments with multiple random seeds on representative base-scale checkpoints.

Our results show that SFT comparisons can depend on the specific pretrained instance. Across tasks, switching checkpoints can change the relative ordering of SFT methods, including cases where architecture and model size are identical and only the pretraining data differs. Multi-seed repetitions further reveal bidirectional transitions between fully activated and under-activated outcomes for the same checkpoint and configuration, which can both expose and mask row-wise top-group differences. Overall, these findings suggest that reported improvements from SFT variants should be interpreted with caution: observed gains may reflect instance- and seed-dependent activation behavior, rather than a universally higher attainable performance ceiling.

\vspace{5mm}

This work
makes two contributions.
\begin{itemize}
\item We provide systematic empirical evidence that the identity of the best-performing (or top-group) SFT configuration can vary across pretrained instances, even when the backbone architecture and the model size are fixed.
\item We introduce the perspective of \textit{elicitation match} to interpret apparent SFT superiority as differences in activation reliability across pretrained instances, explaining why top-group results on a single checkpoint may not transfer consistently.
\end{itemize}

\section{Problem Setup}
\label{sec:method}

\subsection{Problem Statement}
We study \emph{pretraining-dependent instability} in choosing the top SFT configuration for downstream classification: swapping the pretrained SSL checkpoint (within the same backbone family and scale) can change which SFT configuration is best, or statistically indistinguishable from the best, with all other settings fixed.

\subsection{Formal Setup}
Let $M$ be a pretrained SSL checkpoint, $T$ a downstream task, and $s$ a random seed. Each SFT configuration $a\in\mathcal{A}$ yields
\begin{equation}
y=f(a,M,T,s),
\end{equation}
where $y$ is the task metric reported by the evaluator.

\subsection{SFT Configuration Space}
We parameterize configurations by \texttt{FEATURE\_MODE} $m_f\in\{1,2,3\}$ and \texttt{FREEZE\_MODE} $m_z\in\{0,1,2\}$, denoted $a=(m_f,m_z)$. We evaluate
\[
\mathcal{A}=\{(1,0),(1,1),(1,2),(2,0),(2,1),(2,2),(3,0),(3,1)\}.
\]

\subsection{Top-Group Instability Across Checkpoints}
For fixed $(M,T,s)$, let
\begin{equation}
\begin{aligned}
y^\star(M,T,s) &= \max_{a\in\mathcal{A}} f(a,M,T,s),\\
a^\star &= \arg\max_{a\in\mathcal{A}} f(a,M,T,s).
\end{aligned}
\end{equation}
Define the \emph{top group} as configurations not significantly worse than $a^\star$ under the paired exact McNemar test at level $\alpha$:
\begin{equation}
\mathcal{G}(M,T,s;\alpha)=\{a\in\mathcal{A}:\; a \nsucc a^\star \text{ at level }\alpha\}.
\end{equation}
We can suspect \emph{top-group instability} occurs between checkpoints $M$ and $M'$ if
\begin{equation}
\mathcal{G}(M,T,s;\alpha)\neq \mathcal{G}(M',T,s;\alpha).
\end{equation}


\begin{table*}[t]
\centering
\renewcommand{\arraystretch}{1.06}
\caption{Test accuracy computed from paired prediction and truth files for all experiments with seed 1337. Values are bold when they are not significantly different from the row maximum under a two-sided exact McNemar test at 95\% confidence (alpha=0.05); small values are marked as convergence anomalies with an asterisk (*).}\label{tab:main_results}
\adjustbox{max width=\textwidth}{
\begin{tabular}{llllllllll}
\toprule
Task & Model & $F1$-$Z0$ & $F1$-$Z1$ & $F1$-$Z2$ & $F2$-$Z0$ & $F2$-$Z1$ & $F2$-$Z2$ & $F3$-$Z0$ & $F3$-$Z1$ \\
\midrule
\multirow{9}{*}{\textit{SID}} & wav2vec2-base-960h & 0.6590 & 0.7057 & \textbf{0.7833} & \textit{0.0006}* & \textit{0.0006}* & \textit{0.0006}* & 0.6590 & 0.7036 \\
 & wav2vec2-base-100k-voxpopuli & 0.6343 & 0.6769 & \textbf{0.7650} & \textit{0.0006}* & \textit{0.0006}* & \textit{0.0006}* & 0.6343 & 0.6769 \\
 & wav2vec2-large-960h & 0.4156 & 0.7229 & \textbf{0.8294} & \textit{0.0006}* & \textit{0.0006}* & \textit{0.0006}* & 0.4156 & 0.7229 \\
 & wav2vec2-large-100k-voxpopuli & \textbf{0.9066} & \textbf{0.9058} & \textbf{0.9015} & 0.8980 & 0.8811 & \textbf{0.9051} & 0.8981 & \textbf{0.9058} \\
 & hubert-base-ls960 & 0.6834 & 0.7335 & 0.8353 & \textit{0.0006}* & \textbf{0.8753} & \textit{0.0006}* & 0.6834 & 0.7294 \\
 & hubert-large-ll60k & 0.9173 & \textbf{0.9387} & \textbf{0.9386} & 0.9034 & \textbf{0.9335} & 0.9167 & 0.9173 & \textbf{0.9387} \\
 & wavlm-base & \textit{0.1139}* & \textit{0.1977}* & 0.5027 & \textbf{0.7237} & \textbf{0.7148} & \textit{0.0006}* & \textit{0.1190}* & \textit{0.2462}* \\
 & wavlm-base-plus & \textit{0.0934}* & \textit{0.1645}* & 0.5867 & \textit{0.0006}* & \textbf{0.6376} & \textit{0.0006}* & \textit{0.0851}* & \textit{0.1857}* \\
 & wavlm-large & 0.9100 & \textbf{0.9161} & \textbf{0.9230} & 0.9018 & 0.8821 & \textbf{0.9189} & \textbf{0.9213} & 0.9114 \\
\midrule
\multirow{9}{*}{\textit{IC}} & wav2vec2-base-960h & 0.9902 & 0.9939 & \textbf{0.9963} & \textbf{0.9950} & \textbf{0.9955} & \textit{0.1044}* & 0.9902 & 0.9939 \\
 & wav2vec2-base-100k-voxpopuli & 0.9818 & 0.9913 & \textbf{0.9947} & \textit{0.1044}* & \textbf{0.9942} & \textbf{0.9924} & 0.9818 & 0.9913 \\
 & wav2vec2-large-960h & 0.9881 & \textbf{0.9934} & \textbf{0.9950} & \textit{0.1044}* & \textit{0.0910}* & 0.9842 & 0.9881 & \textbf{0.9934} \\
 & wav2vec2-large-100k-voxpopuli & \textbf{0.9968} & \textbf{0.9955} & \textbf{0.9953} & \textbf{0.9966} & \textbf{0.9963} & 0.9945 & \textbf{0.9968} & \textbf{0.9955} \\
 & hubert-base-ls960 & 0.9831 & 0.9902 & \textbf{0.9958} & \textbf{0.9953} & \textbf{0.9960} & \textit{0.1044}* & 0.9831 & 0.9902 \\
 & hubert-large-ll60k & \textbf{0.9955} & \textbf{0.9958} & \textbf{0.9971} & 0.9947 & \textbf{0.9974} & 0.9953 & \textbf{0.9955} & \textbf{0.9958} \\
 & wavlm-base & 0.9818 & 0.9892 & \textbf{0.9963} & \textbf{0.9958} & \textbf{0.9963} & \textit{0.1044}* & 0.9868 & 0.9900 \\
 & wavlm-base-plus & 0.9900 & 0.9929 & \textbf{0.9955} & \textbf{0.9966} & \textbf{0.9966} & 0.9852 & 0.9908 & 0.9908 \\
 & wavlm-large & 0.9958 & \textbf{0.9974} & \textbf{0.9963} & \textbf{0.9963} & \textbf{0.9960} & 0.9953 & \textbf{0.9963} & 0.9939 \\
\midrule
\multirow{9}{*}{\textit{ER}} & wav2vec2-base-960h & 0.5539 & 0.6203 & \textbf{0.7005} & 0.6747 & \textbf{0.7060} & 0.4673 & 0.5539 & 0.6203 \\
 & wav2vec2-base-100k-voxpopuli & 0.5687 & 0.6516 & \textbf{0.6931} & 0.6562 & \textbf{0.6691} & 0.6111 & 0.5687 & 0.6516 \\
 & wav2vec2-large-960h & 0.5548 & \textbf{0.6258} & \textbf{0.6369} & \textit{0.3539}* & \textit{0.3539}* & \textit{0.3539}* & 0.5548 & \textbf{0.6258} \\
 & wav2vec2-large-100k-voxpopuli & 0.6811 & 0.6719 & \textbf{0.7244} & 0.6995 & \textbf{0.7253} & \textbf{0.7032} & 0.6811 & 0.6719 \\
 & hubert-base-ls960 & 0.6608 & 0.6525 & \textbf{0.6765} & 0.6581 & \textbf{0.7032} & \textbf{0.6747} & 0.6608 & 0.6525 \\
 & hubert-large-ll60k & \textbf{0.7134} & \textbf{0.7235} & \textbf{0.7272} & \textbf{0.7161} & \textbf{0.7115} & \textbf{0.7244} & \textbf{0.7134} & \textbf{0.7235} \\
 & wavlm-base & 0.5410 & 0.6018 & \textbf{0.6488} & \textbf{0.6664} & \textbf{0.6581} & 0.6101 & 0.5641 & 0.6065 \\
 & wavlm-base-plus & 0.4922 & 0.5410 & 0.6535 & 0.6406 & \textbf{0.6912} & 0.6359 & 0.6018 & 0.5843 \\
 & wavlm-large & 0.7355 & 0.7272 & 0.7014 & \textbf{0.7770} & 0.7493 & 0.7521 & 0.7456 & 0.7456 \\
\bottomrule
\end{tabular}%
}
\end{table*}

\section{Experiments}

\label{sec:exp}

\subsection{Datasets}
\label{sec:exp_tasks}
We evaluate downstream speech classification on three tasks from the SUPERB benchmark:
intent classification (IC), emotion recognition (ER), and speaker identification (SID).
We follow the official SUPERB data splits and evaluation protocols for each task.
Unless otherwise stated, we use the default evaluation metrics specified by SUPERB for the corresponding task.

\subsection{Backbones}
\label{sec:exp_backbones}
We consider nine widely used self-supervised speech checkpoints spanning three model families:
wav2vec~2.0, HuBERT, and WavLM.
Table~\ref{tab:ssl_models} summarizes each checkpoint, including parameter count and pretraining data.
The table includes same-architecture checkpoints pretrained on different data to enable controlled analysis of
pretraining-instance dependence.

\subsection{Configurations}
\label{sec:exp_configs}
Eight SFT configurations are defined by
\texttt{FEATURE\_MODE}$\in\{1,2,3\}$ and
\texttt{FREEZE\_MODE}$\in\{0,1,2\}$.
Each configuration is denoted as $F m_f$-$Z m_z$
(e.g., $F1$-$Z2$ denotes \texttt{FEATURE\_MODE}=1
and \texttt{FREEZE\_MODE}=2).

\noindent\textbf{Feature modes.}
\texttt{FEATURE\_MODE}=1 uses the final-layer \texttt{last\_hidden\_state} as the supervised representation.
\texttt{FEATURE\_MODE}=2 uses a single selected intermediate layer; we fix the selected layer to the fourth layer from the end, i.e., $\ell^\star=L-3$.
\texttt{FEATURE\_MODE}=3 uses a weighted-sum fusion over transformer layers as the supervised representation.

\noindent\textbf{Freeze modes.}
\texttt{FREEZE\_MODE}=0 fine-tunes the entire pretrained network together with the task head.
\texttt{FREEZE\_MODE}=1 freezes only the convolutional front-end and fine-tunes all transformer layers plus the task head.
\texttt{FREEZE\_MODE}=2 freezes the convolutional front-end and the first $N$ transformer layers, fine-tuning the rest and the task head ($N=4$ in all runs).

\vspace{-2mm}

\subsection{Training Setup}
\label{sec:exp_setup}

\vspace{-1mm}
Our training recipe follows the official SUPERB downstream configurations for IC, ER, and SID as closely as possible. All downstream fine-tuning and evaluation use the official SUPERB training pipeline and built-in evaluator, with identical scripts across runs; only the pretrained checkpoint and random seed are varied. For each task, we keep the entire pipeline fixed across all SSL checkpoints and SFT configurations, including data preprocessing, batching, and validation-based checkpoint selection. Task-specific hyperparameters (optimizer, learning-rate schedule, batch size, and training duration) are taken directly from the corresponding SUPERB configuration files and held constant within each task. Each experiment is run on a single NVIDIA H20 GPU, totaling approximately 10,000 GPU-hours across all runs.

\vspace{-2mm}

\subsection{Seed Protocol}
\label{sec:exp_seed}

\vspace{-1mm}

We use a two-level seed design.
For the full experimental matrix covering all nine SSL checkpoints, we run each (checkpoint, task, SFT configuration) setting with a fixed default seed of \texttt{1337}.
To characterize run-level stochasticity, we additionally perform multi-seed repetitions on three representative base-scale checkpoints:
\texttt{\url{facebook/wav2vec2-base-960h}},
\texttt{facebook/hubert-base-ls960}, and
\texttt{microsoft/wavlm-base}.
For these three checkpoints, we repeat every (task, SFT configuration) setting with two extra seeds, \texttt{2048} and \texttt{7395}, resulting in three seeds in total (\texttt{1337}, \texttt{2048}, \texttt{7395}) for each of the corresponding runs.

\subsection{Evaluation Outputs and Analysis Targets}
\label{sec:exp_targets}
For each run, we report the task-defined metric using the official SUPERB evaluation scripts.
The experimental matrix produces scores indexed by (checkpoint, task, SFT configuration, seed).
All result aggregation, top-group identification via paired exact McNemar tests, and checkpoint- and seed-sensitivity analyses (including under-activation and pairwise sign flips) are presented in Section~\ref{sec:results}.

\begin{table*}[t]
\centering
\renewcommand{\arraystretch}{1.0}
\caption{Test accuracy computed from paired prediction and truth files for experiments with seeds 2048 and 7395, plus seed 1337 for hubert-base-ls960, wav2vec2-base-960h, and wavlm-base, ordered to match the seed-1337 task and model layout. Values are bold when they are not significantly different from the row maximum under a two-sided exact McNemar test at 95\% confidence (alpha=0.05); small values are marked as convergence anomalies with an asterisk (*).}\label{tab:seed_results}
\adjustbox{max width=\textwidth}{
\begin{tabular}{lllllllllll}
\toprule
Task & Model & Seed & $F1$-$Z0$ & $F1$-$Z1$ & $F1$-$Z2$ & $F2$-$Z0$ & $F2$-$Z1$ & $F2$-$Z2$ & $F3$-$Z0$ & $F3$-$Z1$ \\
\midrule
\multirow{9}{*}{\textit{SID}} & wav2vec2-base-960h & 1337 & 0.6590 & 0.7057 & \textbf{0.7833} & \textit{0.0006}* & \textit{0.0006}* & \textit{0.0006}* & 0.6590 & 0.7036 \\
 &  & 2048 & 0.6839 & 0.7292 & \textbf{0.7838} & \textit{0.0006}* & 0.6236 & \textit{0.0006}* & 0.6839 & 0.7427 \\
 &  & 7395 & 0.6722 & 0.7061 & \textbf{0.7738} & \textit{0.0006}* & \textit{0.0006}* & \textit{0.0006}* & 0.6722 & 0.7061 \\
 & hubert-base-ls960 & 1337 & 0.6834 & 0.7335 & 0.8353 & \textit{0.0006}* & \textbf{0.8753} & \textit{0.0006}* & 0.6834 & 0.7294 \\
 &  & 2048 & 0.6759 & 0.6947 & 0.8313 & \textit{0.0006}* & \textbf{0.8670} & \textit{0.0007}* & 0.6759 & 0.6947 \\
 &  & 7395 & 0.6798 & 0.7237 & 0.8474 & 0.8302 & \textbf{0.8738} & \textit{0.0006}* & 0.6798 & 0.7237 \\
 & wavlm-base & 1337 & \textit{0.1139}* & \textit{0.1977}* & 0.5027 & \textbf{0.7237} & \textbf{0.7148} & \textit{0.0006}* & \textit{0.1190}* & \textit{0.2462}* \\
 &  & 2048 & \textit{0.1608}* & \textit{0.2880}* & 0.4585 & \textit{0.0006}* & \textbf{0.6931} & \textit{0.0006}* & \textit{0.0699}* & \textit{0.2624}* \\
 &  & 7395 & \textit{0.1345}* & \textit{0.2082}* & 0.4620 & \textbf{0.7101} & 0.6560 & \textit{0.0006}* & \textit{0.2053}* & \textit{0.2718}* \\
\midrule
\multirow{9}{*}{\textit{IC}} & wav2vec2-base-960h & 1337 & 0.9902 & 0.9939 & \textbf{0.9963} & \textbf{0.9950} & \textbf{0.9955} & \textit{0.1044}* & 0.9902 & 0.9939 \\
 &  & 2048 & 0.9852 & \textbf{0.9937} & \textbf{0.9955} & \textbf{0.9955} & \textbf{0.9955} & \textit{0.1044}* & 0.9852 & \textbf{0.9937} \\
 &  & 7395 & 0.9884 & \textbf{0.9937} & \textbf{0.9955} & \textit{0.1044}* & \textbf{0.9958} & \textbf{0.9945} & 0.9884 & \textbf{0.9937} \\
 & hubert-base-ls960 & 1337 & 0.9831 & 0.9902 & \textbf{0.9958} & \textbf{0.9953} & \textbf{0.9960} & \textit{0.1044}* & 0.9831 & 0.9902 \\
 &  & 2048 & 0.9871 & \textbf{0.9931} & \textbf{0.9960} & \textit{0.0464}* & \textbf{0.9960} & \textit{0.1044}* & 0.9871 & \textbf{0.9931} \\
 &  & 7395 & 0.9831 & 0.9934 & \textbf{0.9963} & \textit{0.0464}* & \textbf{0.9958} & \textit{0.0464}* & 0.9831 & 0.9934 \\
 & wavlm-base & 1337 & 0.9818 & 0.9892 & \textbf{0.9963} & \textbf{0.9958} & \textbf{0.9963} & \textit{0.1044}* & 0.9868 & 0.9900 \\
 &  & 2048 & 0.9839 & 0.9821 & \textbf{0.9960} & 0.9924 & \textbf{0.9955} & \textit{0.1044}* & 0.9829 & 0.9871 \\
 &  & 7395 & 0.9855 & 0.9900 & \textbf{0.9958} & \textit{0.0000}* & \textbf{0.9953} & \textit{0.1044}* & 0.9831 & 0.9889 \\
\midrule
\multirow{9}{*}{\textit{ER}} & wav2vec2-base-960h & 1337 & 0.5539 & 0.6203 & \textbf{0.7005} & 0.6747 & \textbf{0.7060} & 0.4673 & 0.5539 & 0.6203 \\
 &  & 2048 & 0.5585 & 0.6175 & \textbf{0.6820} & \textit{0.3548}* & \textbf{0.6940} & \textit{0.2562}* & 0.5585 & 0.6175 \\
 &  & 7395 & 0.5585 & 0.6424 & \textbf{0.6949} & 0.6590 & \textbf{0.7134} & 0.6175 & 0.5585 & 0.6424 \\
 & hubert-base-ls960 & 1337 & 0.6608 & 0.6525 & \textbf{0.6765} & 0.6581 & \textbf{0.7032} & \textbf{0.6747} & 0.6608 & 0.6525 \\
 &  & 2048 & 0.6350 & \textbf{0.6654} & \textbf{0.6562} & \textit{0.3539}* & \textbf{0.6728} & \textbf{0.6654} & 0.6350 & \textbf{0.6654} \\
 &  & 7395 & 0.6258 & 0.6488 & \textbf{0.6857} & \textbf{0.6866} & \textbf{0.6793} & \textbf{0.6719} & 0.6258 & 0.6488 \\
 & wavlm-base & 1337 & 0.5410 & 0.6018 & \textbf{0.6488} & \textbf{0.6664} & \textbf{0.6581} & 0.6101 & 0.5641 & 0.6065 \\
 &  & 2048 & 0.5558 & 0.5991 & 0.6092 & 0.6719 & \textbf{0.7060} & 0.4931 & 0.4986 & 0.5364 \\
 &  & 7395 & 0.6212 & 0.6055 & 0.6636 & 0.6645 & \textbf{0.7217} & 0.6194 & 0.5078 & 0.6138 \\
\bottomrule
\end{tabular}%
}
\end{table*}

\section{Results}

\label{sec:results}

Each row in Tables~\ref{tab:main_results} and~\ref{tab:seed_results} corresponds to a fixed downstream task and a fixed pretrained checkpoint, under which we compare the eight SFT configurations in Section~\ref{sec:exp}. We mark systems that are not significantly worse than the row-wise best as the \emph{top group}, i.e., statistically indistinguishable from the best under the paired exact McNemar test \cite{McNemar_1947} at $\alpha=0.05$.

\subsection{Why the differences are not ceiling improvements}
At a glance, bold entries are not uniformly distributed across columns, and some SFT configurations enter the top group more frequently across pretrained checkpoints. A naive interpretation is that these configurations have an intrinsically higher attainable performance ceiling. However, several observations are inconsistent with a ceiling-improvement explanation.

First, configurations that frequently enter the top group can still exhibit \emph{under-activation} on certain checkpoints, producing abnormally low performance on the same task. If the advantage primarily came from a systematically higher ceiling, one would expect the relative ordering to be more consistent and the ``best'' configurations to rarely collapse. Instead, extreme failures among apparent winners indicate that the observed gaps are not simply explained by different upper bounds.

Second, Table~\ref{tab:seed_results} shows that the effect is, to some extent, seed-dependent and \emph{bidirectional}. For the same task, pretrained checkpoint, and SFT configuration, changing only the seed can occasionally turn an under-activated run into a fully activated one that reaches the row-wise top group statistically indistinguishable from the best. Conversely, a configuration that is in the top group under one seed can become under-activated (or markedly degraded) under another seed, without changing data, hyperparameters, or training mode. Such bidirectional transitions are difficult to reconcile with a stable ceiling-improvement story, but are consistent with sensitivity to the optimization trajectory.

Finally, for the \texttt{hubert-large-ll60k} checkpoint on the \textit{ER} task, all eight configurations achieve statistically indistinguishable performance, placing the entire row in the top group. This limiting case indicates that, for some pretrained instances under our protocol, the choice among these SFT configurations has limited marginal effect on the final metric, further weakening the claim that certain configurations provide a generally higher ceiling.

\subsection{Why the differences reflect activation matching}
Taken together, the results are better explained by an \emph{activation matching} perspective: different SFT configurations differ primarily in how reliably they can \emph{activate} the latent capability of a given pretrained checkpoint under a particular training stochasticity, rather than in how high they can ultimately go.

The non-uniform bold-entry distribution then reflects differences in \emph{activation success rate}. A configuration that appears in the top group more often is one that more frequently reaches a high-performing solution region for many checkpoints under the default recipe. Under-activation corresponds to a failure to enter that region for a specific checkpoint--seed combination, even for configurations that otherwise succeed often.

The seed results provide direct support for this view. The fact that activation can be flipped from failure to success (and vice versa) by changing only the seed suggests that the main variability lies in whether optimization finds a suitable basin that aligns the SFT configuration with the pretrained representation. When activation succeeds, multiple configurations can reach comparable top-group performance for the same checkpoint, implying that the primary distinction is \emph{how easy it is to get there} rather than a systematically higher attainable ceiling.

Notably, the abnormally low scores marked as convergence anomalies (*) do not stem from accidental training interruptions: all runs complete the full training pipeline under the same configuration. This further confirms that poor performance primarily reflects a failure of the SFT procedure to successfully activate the pretrained model's usable capacity.

\section{Conclusion}


Across nine SSL checkpoints and eight SFT configurations, we show that selecting the statistically indistinguishable top-group SFT recipe is pretrained-instance dependent. Even with fixed architecture and scale, the best-performing configuration can change across checkpoints, and pairwise orderings may flip as a secondary effect. Multi-seed results further reveal this fragile activation, where the same recipe can alternate between fully activated and under-activated outcomes due to optimization stochasticity. Overall, apparent SFT gains often reflect instance and seed chdependent activation reliability rather than universally better recipes, motivating evaluation across multiple checkpoints and seeds.

\section{Acknowledgments}
This work was supported by JST BOOST, Grant Number JPMJBS2407.

\section{Generative AI Use Disclosure}
Generative AI tools (Gemini and ChatGPT) were used for language editing and improving the phrasing of this manuscript.

\bibliographystyle{IEEEtran}

\bibliography{mybib}

@article{baevski2020wav2vec,
  title={wav2vec 2.0: A framework for self-supervised learning of speech representations},
  author={Baevski, Alexei and Zhou, Yuhao and Mohamed, Abdelrahman and Auli, Michael},
  journal={Advances in neural information processing systems},
  volume={33},
  pages={12449--12460},
  year={2020}
}

@article{hsu2021hubert,
  title={Hubert: Self-supervised speech representation learning by masked prediction of hidden units},
  author={Hsu, Wei-Ning and Bolte, Benjamin and Tsai, Yao-Hung Hubert and Lakhotia, Kushal and Salakhutdinov, Ruslan and Mohamed, Abdelrahman},
  journal={IEEE/ACM transactions on audio, speech, and language processing},
  volume={29},
  pages={3451--3460},
  year={2021},
  publisher={IEEE}
}

@article{chen2022wavlm,
  title={Wavlm: Large-scale self-supervised pre-training for full stack speech processing},
  author={Chen, Sanyuan and Wang, Chengyi and Chen, Zhengyang and Wu, Yu and Liu, Shujie and Chen, Zhuo and Li, Jinyu and Kanda, Naoyuki and Yoshioka, Takuya and Xiao, Xiong and others},
  journal={IEEE Journal of Selected Topics in Signal Processing},
  volume={16},
  number={6},
  pages={1505--1518},
  year={2022},
  publisher={IEEE}
}

@article{yang2021superb,
  title={SUPERB: Speech Processing Universal PERformance Benchmark},
  author={Yang, Shu-wen and Chi, Po-Han and Chuang, Yung-Sung and Lai, Cheng-I Jeff and Lakhotia, Kushal and Lin, Yist Y and Liu, Andy T and Shi, Jiatong and Chang, Xuankai and Lin, Guan-Ting and others},
  journal={Interspeech},
  year={2021},
  publisher={ISCA}
}

@inproceedings{gao2023two,
  title={Two-stage Finetuning of Wav2vec 2.0 for Speech Emotion Recognition with ASR and Gender Pretraining},
  author={Gao, Yuan and Chu, Chenhui and Kawahara, Tatsuya},
  booktitle={Proc. Interspeech},
  pages={3637--3641},
  year={2023}
}

@inproceedings{gao2024enhancing,
  title={Enhancing two-stage finetuning for speech emotion recognition using adapters},
  author={Gao, Yuan and Shi, Hao and Chu, Chenhui and Kawahara, Tatsuya},
  booktitle={Proc. IEEE International Conference on Acoustics, Speech and Signal Processing (ICASSP)},
  pages={11316--11320},
  year={2024},
}

@INPROCEEDINGS{10446041,
  author={Zhou, Wangjin and Yang, Zhengdong and Chu, Chenhui and Li, Sheng and Dabre, Raj and Zhao, Yi and Tatsuya, Kawahara},
  booktitle={Proc. IEEE International Conference on Acoustics, Speech and Signal Processing (ICASSP)}, 
  title={MOS-FAD: Improving Fake Audio Detection Via Automatic Mean Opinion Score Prediction}, 
  year={2024},
  volume={},
  number={},
  pages={876-880},
  doi={10.1109/ICASSP48485.2024.10446041}}

@INPROCEEDINGS{10848773,
  author={Shi, Xiaohan and Gao, Yuan and He, Jiajun and Mi, Jinyi and Li, Xingfeng and Toda, Tomoki},
  booktitle={Proc. Asia Pacific Signal and Information Processing Association Annual Summit and Conference (APSIPA ASC)}, 
  title={A Study on Multimodal Fusion and Layer Adapter in Emotion Recognition}, 
  year={2024},
  volume={},
  number={},
  pages={1-6},
  doi={10.1109/APSIPAASC63619.2025.10848773}}

@INPROCEEDINGS{10889634,
  author={Zhao, Jiahui and Shi, Hao and Cui, Chenrui and Wang, Tianrui and Liu, Hexin and Ni, Zhaoheng and Ye, Lingxuan and Wang, Longbiao},
  booktitle={Proc. IEEE International Conference on Acoustics, Speech and Signal Processing (ICASSP)}, 
  title={Adapting Whisper for Code-Switching through Encoding Refining and Language-Aware Decoding}, 
  year={2025},
  volume={},
  number={},
  pages={1-5},
  doi={10.1109/ICASSP49660.2025.10889634}}

@ARTICLE{11353442,
  author={Shi, Xiaohan and He, Jiajun and Li, Xingfeng and Toda, Tomoki},
  journal={IEEE Transactions on Audio, Speech and Language Processing}, 
  title={A Comprehensive Study on the Effectiveness of ASR Representations for Noise-Robust Speech Emotion Recognition}, 
  year={2026},
  volume={34},
  number={},
  pages={707-722},
  doi={10.1109/TASLPRO.2026.3654273}}

@article{Zhang2025SONARSC,
  title={{SONAR}: Self-Distilled Continual Pre-training for Domain Adaptive Audio Representation},
  author={Yizhou Zhang and Yuan Gao and Wangjin Zhou and Zichen Yuan and Keisuke Imoto and Tatsuya Kawahara},
  journal={ArXiv},
  year={2025},
  volume={abs/2509.15703}
}

@inproceedings{sang2024efficient,
  title={Efficient adapter tuning of pre-trained speech models for automatic speaker verification},
  author={Sang, Mufan and Hansen, John HL},
  booktitle={Proc. IEEE International Conference on Acoustics, Speech and Signal Processing (ICASSP)},
  pages={12131--12135},
  year={2024},
  organization={IEEE}
}

@inproceedings{merchant-etal-2020-happens,
    title = "What Happens To {BERT} Embeddings During Fine-tuning?",
    author = "Merchant, Amil  and
      Rahimtoroghi, Elahe  and
      Pavlick, Ellie  and
      Tenney, Ian",
    editor = "Alishahi, Afra  and
      Belinkov, Yonatan  and
      Chrupa{\l}a, Grzegorz  and
      Hupkes, Dieuwke  and
      Pinter, Yuval  and
      Sajjad, Hassan",
    booktitle = "Proceedings of the Third BlackboxNLP Workshop on Analyzing and Interpreting Neural Networks for NLP",
    month = nov,
    year = "2020",
    publisher = "Association for Computational Linguistics",
    pages = "33--44"

}

@article{Zhang2025OnTI,
  title={On the Interplay of Pre-Training, Mid-Training, and RL on Reasoning Language Models},
  author={Charlie Zhang and Graham Neubig and Xiang Yue},
  journal={ArXiv},
  year={2025},
  volume={abs/2512.07783},
}

@inproceedings{Tian2020ImprovingES,
  title={Improving End-to-End Speech-to-Intent Classification with Reptile},
  author={Yusheng Tian and Philip John Gorinski},
  booktitle={Proc. Interspeech},
  year={2020},
}

@INPROCEEDINGS{9413388,
  author={Sharma, Bidisha and Madhavi, Maulik and Li, Haizhou},
  booktitle={Proc. IEEE International Conference on Acoustics, Speech and Signal Processing (ICASSP)}, 
  title={Leveraging Acoustic and Linguistic Embeddings from Pretrained Speech and Language Models for Intent Classification}, 
  year={2021},
  volume={},
  number={},
  pages={7498-7502},
  doi={10.1109/ICASSP39728.2021.9413388}}

@ARTICLE{11270007,
author={Gao, Yuan and Shi, Hao and Fu, Yahui and Chu, Chenhui and Kawahara, Tatsuya},
journal={ IEEE Transactions on Affective Computing },
title={{ Bridging Speech Emotion Recognition and Personality: Dataset and Temporal Interaction Condition Network }},
year={2026},
volume={17},
number={01},
ISSN={1949-3045},
pages={829-840},
address={Los Alamitos, CA, USA},
month=jan}

@INPROCEEDINGS{9747870,
  author={Morais, Edmilson and Hoory, Ron and Zhu, Weizhong and Gat, Itai and Damasceno, Matheus and Aronowitz, Hagai},
  booktitle={Proc. IEEE International Conference on Acoustics, Speech and Signal Processing (ICASSP)}, 
  title={Speech Emotion Recognition Using Self-Supervised Features}, 
  year={2022},
  volume={},
  number={},
  pages={6922-6926},
  doi={10.1109/ICASSP43922.2022.9747870}}

@INPROCEEDINGS{10094954,
  author={Han, Bing and Chen, Zhengyang and Qian, Yanmin},
  booktitle={Proc. IEEE International Conference on Acoustics, Speech and Signal Processing (ICASSP)}, 
  title={Exploring Binary Classification Loss for Speaker Verification}, 
  year={2023},
  volume={},
  number={},
  pages={1-5}}

@ARTICLE{11075516,
  author={Chen, Zhiyong and Wu, Shuhang and Li, Xinnuo and Ai, Zhiqi and Xu, Shugong},
  journal={IEEE Transactions on Audio, Speech and Language Processing}, 
  title={Open-Set Speaker Identification Through Efficient Few-Shot Tuning With Speaker Reciprocal Points and Unknown Samples}, 
  year={2025},
  volume={33},
  number={},
  pages={3347-3362},
  doi={10.1109/TASLPRO.2025.3587591}}

@article{McNemar_1947, title={Note on the Sampling Error of the Difference Between Correlated Proportions or Percentages}, volume={12}, DOI={10.1007/BF02295996}, number={2}, journal={Psychometrika}, author={McNemar, Quinn}, year={1947}, pages={153–157}}

\end{document}